**CABS-flex standalone 3: an open command-line platform for protein flexibility simulation, peptide structure modeling, and protein-peptide docking**

Chandran Nithin, Karol Wroblewski, Piotr Szukało, Ayomide Fasemire, Aleksander Kuriata, Mateusz Kurcinski, Andrzej Kolinski, Sebastian Kmiecik*

Faculty of Chemistry, Biological and Chemical Research Centre, University of Warsaw, Warsaw, Poland

**Summary:** CABS-flex standalone 3 is an open command-line platform for fast CABS-based coarse-grained modeling of protein flexibility, peptide structures, and global or information-guided protein-peptide docking, coupled with all-atom reconstruction and analysis. The package builds on the established CABS-flex and CABS-dock ecosystem, widely used in structural bioinformatics for protein flexibility simulations and flexible protein–peptide docking. It provides a Python 3 implementation that brings together previous standalone functionality with recent developments in protein flexibility simulation, linear and cyclic peptide modeling, extended reporting and visualization, and deep-learning-based all-atom reconstruction with cg2all.

**Availability and Implementation:** CABS-flex standalone 3 is implemented in Python 3 and is freely available as an open-source command-line package. Documentation is available at https://cabsflex.lcbio.pl. Source code is available at https://github.com/LCBio/CABSflex_standalone.

**Contact:** sekmi@chem.uw.edu.pl

**Supplementary information:** Supplementary data, examples, and documentation are available online at https://cabsflex.lcbio.pl.

## 1. Introduction

Protein and peptide conformational flexibility is central to molecular recognition, regulation, conformational switching, and biomolecular interaction mechanisms. Static structural models, including experimentally determined structures and high-confidence predicted models, often represent only one accessible state of a biomolecule, and there is increasing interest in computational approaches that move beyond static structures toward conformational ensembles and protein dynamic conformations (Cui et al. 2025). In practical applications, this creates a need for conformational ensembles, local flexibility profiles, alternative binding-site conformations, peptide structure models, and flexible protein-peptide docking poses. These tasks require methods that can efficiently sample conformational variability while remaining practical for routine use, batch calculations, and reproducible command-line workflows.

The CABS model is a reduced-representation protein model developed for efficient conformational sampling of proteins and peptides (Kolinski 2004, Kmiecik et al. 2016). It has provided the methodological basis for CABS-flex and CABS-dock, two well-established tools for protein flexibility simulations and flexible protein-peptide docking. CABS-flex has been validated against all-atom molecular dynamics simulations and NMR ensembles (Jamroz et al. 2013a, Jamroz et al. 2014), made available through web-server and standalone implementations (Jamroz et al. 2013b, Kuriata et al. 2018, Kurcinski et al. 2019a), and applied to a broad range of structure–dynamics–function studies (Nithin et al. 2024). CABS-dock was introduced as a web server for global flexible protein-peptide docking without prior knowledge of the binding site (Kurcinski et al. 2015), later extended into the CABS-dock standalone package (Kurcinski et al. 2019b), further developed for information-driven docking using contact restraints (Blaszczyk et al. 2019), and applied across multiple protein-peptide docking scenarios (Kurcinski et al. 2020). Together, these tools form a mature CABS-based modeling ecosystem for protein flexibility, peptide conformational sampling and protein-peptide interactions.

Here we present CABS-flex standalone 3, a major open Python 3 implementation that continues a longer line of CABS-based software development. Building on the previous CABS-flex and CABS-dock standalone packages (Kurcinski et al. 2019a,b), it integrates earlier web-server and standalone workflows into a single command-line platform. The new version supports three main workflows: protein flexibility simulation, *de novo* modeling of linear and cyclic peptides, and protein-peptide docking. It connects CABS coarse-grained sampling with restraint-controlled simulations, model selection, clustering, all-atom reconstruction, visualization, and reporting. At the same time, the package gives users more control than is typically available in web-server workflows: customized restraints, reproducible benchmark studies, larger modeling pipelines and automated analyses.

## 2. Workflows and recent methodological updates

As summarized above, CABS-flex standalone 3 builds on the established CABS-flex and CABS-dock ecosystem. The new version updates the previous CABS-flex and CABS-dock standalone packages (Kurcinski et al. 2019a,b) from Python 2 to Python 3 and integrates their workflows into a unified command-line platform with updated reconstruction, reporting, and visualization procedures. A central technical upgrade is the integration of the deep-learning-based cg2all reconstruction method, which improves the local structural quality of all-atom models reconstructed from CABS coarse-grained representations (Heo and Feig 2024). By default, OpenMM-based optimization further refines the reconstructed structures, increasing their usefulness for visual inspection, interpretation of peptide conformations/docking poses, and downstream structural analysis.

CABS-flex standalone 3 provides a unified interface to three workflow families: protein flexibility simulations, peptide modeling, and protein-peptide docking. These workflows share a common modeling core based on CABS coarse-grained simulation, restraint-tuned sampling, filtering, scoring, clustering, all-atom reconstruction, and final output analysis (Fig. 1A).

The protein flexibility workflow is designed for fast simulations of protein conformational flexibility around an input structure. It generates representative near-native models (ten by default), that can be used to analyze local and global flexibility. Building on developments introduced in the CABS-flex 3.0 web server (Wróblewski et al. 2025), this workflow supports new flexibility modes with adjustable restraint schemes, including pLDDT-guided restraints and rigid-like restraint settings for stable, well-folded globular proteins (Wróblewski and Kmiecik 2024).

The peptide modeling workflow supports *de novo* modeling of linear and cyclic peptides using coarse-grained CABS sampling followed by all-atom reconstruction and optimization. It builds on the multiscale protocol for structure prediction of linear, backbone-cyclized, and disulfide-constrained peptides (Badaczewska-Dawid et al. 2024), as well as on functionality introduced in the CABS-flex 3.0 web server (Wróblewski et al. 2025). Cyclic and disulfide-constrained systems are modeled by introducing appropriate distance restraints that close the peptide chain or enforce known covalent constraints.

The protein-peptide docking workflow builds on CABS-dock, an established method for flexible protein-peptide docking described above. It supports global peptide docking without prior knowledge of the binding site, as well as information-guided docking using contact restraints. The peptide is treated as fully flexible, while receptor flexibility can be controlled through near-native restraint settings. Recent extensions include cyclic peptide docking protocols combining CABS-dock sampling with cyclic restraints and refinement procedures (Zalewski et al. 2025).

Across all three workflows, the CABS-flex standalone 3 implementation provides command-line control beyond the web-server interfaces. Users can modify restraint definitions, adjust sampling parameters, control model selection and scoring options, customize clustering and analysis outputs, choose all-atom reconstruction and optimization settings, as well as integrate the workflows into larger automated modeling pipelines. This makes the package suitable for reproducible studies, batch calculations, customized protocols, and downstream structural bioinformatics analyses.

## 3. Implementation and outputs

CABS-flex standalone 3 is implemented in Python 3 and distributed as an open-source command-line package. The software provides workflow-specific commands and options while preserving a shared internal architecture for simulation, model selection, reconstruction, and analysis. Input data includes protein structures, peptide sequences, optional secondary-structure definitions, pLDDT information, and user-defined restraints.

The package generates organized output directories containing coarse-grained trajectories, selected models, clustered representatives, reconstructed all-atom structures, and analysis files. Depending on selected options, users can obtain PDB files, CSV and JSON summaries, RMSF profiles, contact maps, restraint reports, energy and scoring summaries, and files supporting molecular visualization. The software can also generate Jupyter notebooks and HTML reports that summarize key outputs, including interactive structure visualization, RMSF plots, and contact maps.

All-atom reconstruction relies on cg2all-based rebuilding, supplemented by MODELLER-based reconstruction where required (Sali and Blundell 1993). Reconstructed structures may also be further optimized using OpenMM. This flexibility allows users to choose reconstruction and refinement settings tailored to their system and downstream goals, such as visualization, inspecting peptide conformations or docking poses, and preparing structures for further analysis.

The documentation provides ready-to-run examples and visualization instructions for the main workflows, including ensemble visualization, RMSF-based coloring, contact-map analysis, peptide-model inspection, docking-pose visualization, and energy-versus-RMSD plots.

## 4. Example uses

The package documentation provides ready-to-run examples for the three main application areas. Table 1 shows basic commands illustrating the standard syntax for protein flexibility simulation, peptide structure prediction and protein-peptide docking. By default, the workflows generate 10 representative models selected from the sampled conformational space, together with workflow-specific analysis and visualization outputs.

**Table 1. Basic CABS-flex standalone 3 command-line workflows.**

| Workflow | Basic command* | Main output |
|---|---|---|
| Protein flexibility | `CABSflex -i 1hpw` | Protein conformational ensemble; RMSF profile |
| Peptide modeling | `CABSflex --peptide ACDEFGHIKLMNPQRSTVWY` | Predicted peptide structure models |

| Protein-peptide docking | `CABSdock -i 2fvj:A -p HKLVQLLTTT:CHHHHHHHCC` | Predicted protein-peptide docking poses; clustering/scoring summaries |
|---|---|---|

*In these examples, `1hpw` and `2fvj` are PDB identifiers, while `A` indicates the receptor chain used for docking. In the docking command, `HKLVQLLTTT` is the peptide sequence and `CHHHHHHHCC` is an optional peptide secondary-structure string.

In a typical protein flexibility simulation, the user provides an input protein structure or PDB identifier and obtains representative near-native models together with residue-level fluctuation profiles (Fig. 1B). These outputs can be used to identify flexible loops, termini, hinge-like regions, or alternative conformational states. In peptide structure prediction, the user provides a peptide sequence with optional secondary-structure or restraint information, and the software returns representative peptide conformations. In protein-peptide docking, the peptide is sampled flexibly around the receptor or near a contact-guided binding region. The resulting output includes representative docking poses, clustering/scoring summaries, and energy-versus-RMSD plots of the sampled docking landscape (Fig. 1C).

## 5. Discussion

CABS-flex standalone 3 provides a reproducible command-line environment for CABS-based modeling of protein and peptide conformational variability. Its main advantage is the integration of fast coarse-grained sampling with user-controllable restraints, workflow-specific options, all-atom reconstruction, reporting, and visualization. This makes the package suitable for users who need open, scriptable access to CABS-flex/CABS-dock functionality, including batch calculations, custom restraint definitions, reproducible benchmarks, and integration with external structural bioinformatics pipelines. More broadly, the command-line format and editable restraint definitions make the package suitable for developing information-guided modeling protocols that combine CABS sampling with experimental constraints or external structural hypotheses. User-defined distance or contact restraints can be used to incorporate prior information into protein flexibility simulations, peptide modeling or protein-peptide docking, for example, when interpreting experimental data or testing alternative structural scenarios.

The software is not intended to replace all-atom molecular dynamics. CABS simulations generate coarse-grained, pseudo-dynamic conformational ensembles rather than physically time-resolved atomistic trajectories (Kmiecik et al. 2016). Similarly, docking scores should be interpreted as model-selection and prioritization measures rather than rigorous binding free energies. For applications requiring detailed atomic interactions, explicit solvent effects or precise energetics, CABS-flex standalone 3 can be used as an upstream ensemble-generation and pose-generation step followed by all-atom refinement or complementary downstream analyses.

By bringing together established CABS-flex and CABS-dock methods with recent developments in restraint schemes, peptide modeling, contact-guided docking, cyclic peptide docking, report generation, and cg2all-based reconstruction, CABS-flex standalone 3 extends the CABS ecosystem from separate web-server and standalone tools toward a more integrated, scriptable framework for structural bioinformatics.

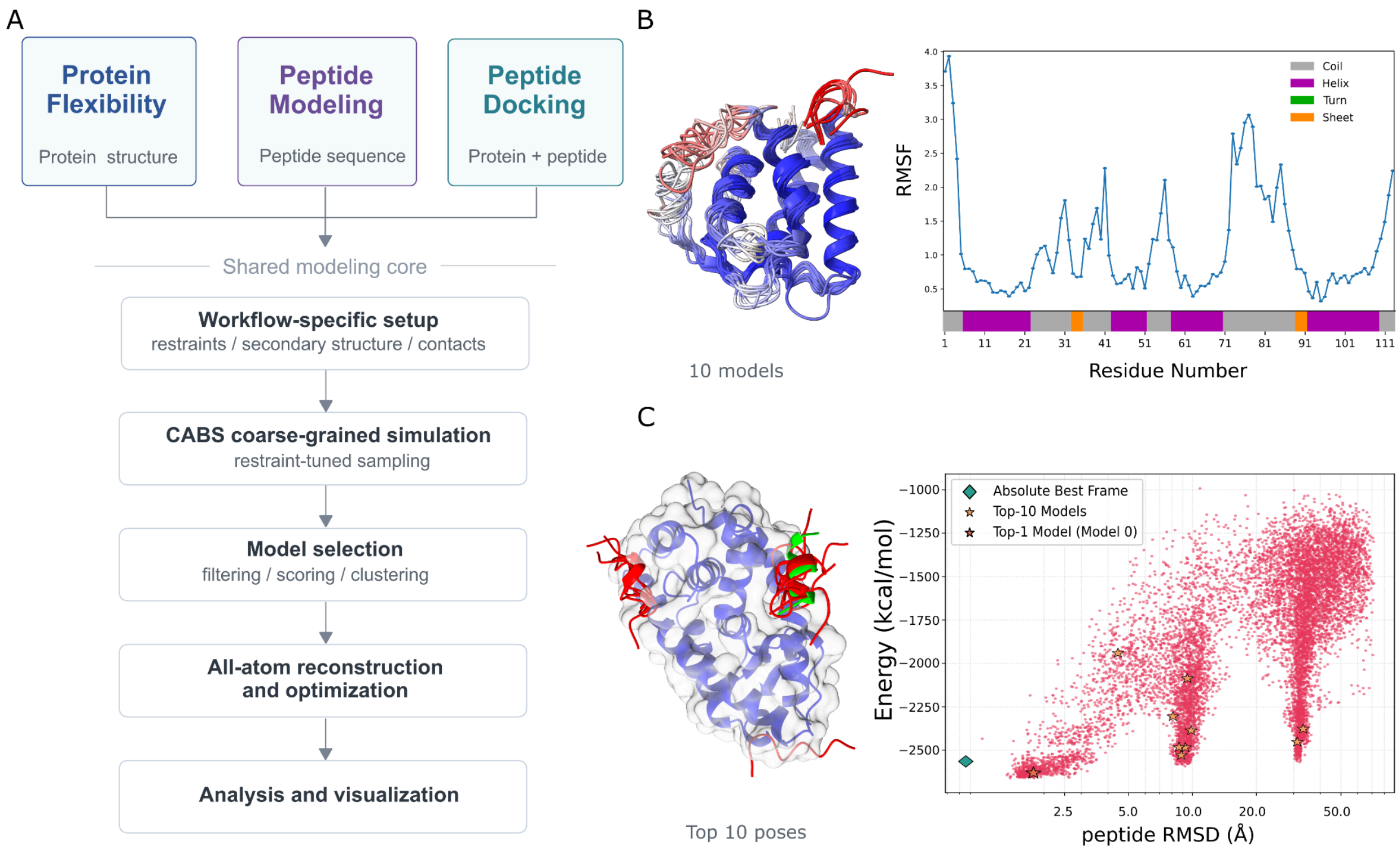

**Figure 1. Overview of CABS-flex standalone 3 workflows and example outputs. A**, three main workflows—protein flexibility, peptide modeling, and protein-peptide docking—converge into a shared modeling core consisting of workflow-specific setup, CABS coarse-grained simulation, model selection, all-atom reconstruction and optimization, and analysis/visualization. **B**, example protein flexibility outputs include a conformational ensemble of 10 representative models and an RMSF profile. **C**, example protein-peptide docking outputs include 10 top docking poses and an energy-versus-RMSD plot summarizing the sampled docking landscape.

## Availability

Documentation and examples are available at https://cabsflex.lcbio.pl. Source code is available at https://github.com/LCBio/CABSflex_standalone.

## Acknowledgments

The authors thank all current and former contributors for the development of the CABS-flex and CABS-dock software ecosystem. In particular, we acknowledge Michal Jamroz, Aleksandra E. Badaczewska-Dawid, Mateusz Zalewski, Jan Kulczycki, Maciej P. Ciemny, Tymoteusz Oleniecki, Michał Kolinski and Rafał Madaj for their contributions to earlier standalone implementations, related tools, documentation and software-development efforts that helped establish the basis for the current CABS-flex standalone 3 release. We acknowledge support from the National Science Centre, Poland (Sheng grant no. 2021/40/Q/NZ2/00078), and the Polish high-performance computing infrastructure PLGrid (HPC centers: ACK Cyfronet AGH and WCSS) for providing computational resources and support under computational grant no. PLG/2026/019105.

## Conflict of interests

None declared